\documentclass[aps,prl,reprint,superscriptaddress,showpacs,nofootinbib]{revtex4-2}
\usepackage{amssymb}
\usepackage{amsmath}
\usepackage{graphicx}
\usepackage{hyperref}
\usepackage{multirow,diagbox,xcolor}

\newcommand{\Pl}{\ell_\mathrm{p}} 

\newcommand{\sgn}{\mathrm{sgn}} 
\newcommand{\hil}{\mathcal{H}} 
\newcommand{\hmu}{\mathcal{H}_{\mathfrak b}}
\newcommand{\htau}{\mathcal{H}_{\mathfrak c}}

\newcommand{\dd}{{\rm d}}
\newcommand{\h}{\mathfrak{h}}
\newcommand{\hbh}{\hil_{\rm BH}}
\renewcommand{\b}{\mathfrak b}
\renewcommand{\c}{\mathfrak c}

\newcommand{\hmuep}{\hmu^{\varepsilon_b}}
\newcommand{\htauep}{\htau^{\varepsilon_c}}

\makeatletter
\newsavebox{\@brx}
\newcommand{\llangle}[1][]{\savebox{\@brx}{\(\m@th{#1\langle}\)}%
  \mathopen{\copy\@brx\kern-0.5\wd\@brx\usebox{\@brx}}}
\newcommand{\rrangle}[1][]{\savebox{\@brx}{\(\m@th{#1\rangle}\)}%
  \mathclose{\copy\@brx\kern-0.5\wd\@brx\usebox{\@brx}}}
\newcommand{\llb}[1][]{\savebox{\@brx}{\(\m@th{#1(}\)}%
  \mathopen{\copy\@brx\kern-0.5\wd\@brx\usebox{\@brx}}}
\newcommand{\rrb}[1][]{\savebox{\@brx}{\(\m@th{#1)}\)}%
  \mathclose{\copy\@brx\kern-0.5\wd\@brx\usebox{\@brx}}}
\makeatother

\newcommand{\hpb}{\hat{p}_b}
\newcommand{\hpc}{\hat{p}_c}
\newcommand{\hp}{\hat{\h}}
\newcommand{\hpm}{\hp^{(m)}}

\begin{document}


\title{Loop quantum Schwarzschild interior and black hole remnant}


\author{Cong Zhang}
\affiliation{Faculty of Physics, University of Warsaw, Pasteura 5, 02-093 Warsaw, Poland}
\author{Yongge Ma}
\thanks{mayg@bnu.edu.cn}
\author{Shupeng Song}
\thanks{songsp@mail.bnu.edu.cn}
\affiliation{Department of Physics, Beijing Normal University, Beijing 100875, China}
\author{Xiangdong Zhang}
\thanks{scxdzhang@scut.edu.cn}
\affiliation{Department of Physics, South China University of Technology, Guangzhou 510641, China}

\date{\today}

\begin{abstract}
The interior of Schwarzschild black hole is quantized by the method of loop quantum gravity. The Hamiltonian constraint is solved and the physical Hilbert space is obtained in the model. The properties of a Dirac observable corresponding to the ADM mass of the Schwarzschild black hole are studied by both analytical and numerical techniques. It turns out that zero is not in the discrete spectrum of this Dirac observable. This supports the existence of a stable remnant after the evaporation of a black hole. Our conclusion is valid for a general class of schemes adopted for loop quantization of the model.  
\end{abstract}

\maketitle

The quantum nature of black hole (BH) is a challenging topic concerning the unification of general relativity, quantum mechanics and statistical mechanics. To study the final state of BH evaporation is relevant to the constituent of dark matter.  According to the analysis of Hawking radiation \cite{hawking1974black}, the primordial mini BHs in the very early universe should be completely evaporated by now. However, if the BH evaporation halted at some stable state, which is called the BH remnant, because of some quantum gravity effect, these remnants would have important cosmological consequences \cite{barrow1992cosmology,carr1994black,dalianis2019primordial}. 
Remarkably, the remnants of the primordial black holes could even comprise the entire dark matter in the universe \cite{barrow1992cosmology,carr1994black}. 
Moreover, the existence of the BH remnant could provide a possible approach to solve the puzzle of information loss \cite{preskill1992black,chen2015black}. Thanks to the remnant, one could argue that the information fallen into a BH with matters could be stored in the remnant after the evaporation. Furthermore, one could also analogize a BH to an atom in quantum mechanics to argue the distortion of the semiclassical Hawking spectrum resulted from the discreteness of the BH mass \cite{bekenstein1997quantum,lochan2016discrete}. 
It was argued that in certain cases, the distortion could be observable even for macroscopic black holes \cite{bekenstein1997quantum}. Although the debates on the quantum nature of BH are crucial and long standing, there is no systematic study by quantum gravity so far to lay a solid theoretical foundation for the arguments.

As a background independent approach to quantum gravity, loop quantum gravity (LQG) has been widely studied in the past 30 yeas \cite{rovelli2004quantum,ashtekar2004background,han2007fundamental,thiemann2008modern,abhay2017loop}. Recent works in the symmetry-reduced models of LQG indicate that the singularity inside a Schwarzschild BH can be resolved by LQG effects \cite{ashtekar2005quantum,modesto2006loop,boehmer2007loop,chiou2008phenomenological,gambini2013loop,corichi2016loop,ashtekar2018quantum,bojowald2018effective,bodendorfer2019effective}. However, unlike the singularity problem which can be discussed with macroscopic BHs, the final of BH evaporation should be purely quantum and thus could not be described by notions of classical geometry. Hence, in order to deal with the issue of BH final state, it is crucial to come up with some new notions and techniques of quantum gravity. The notable theory of LQG may take on the responsibilities. 

In this paper, we address the issue of quantum BH by the symmetry-reduced model of LQG.  By loop quantization of the interior of a Schwarzschild BH, a Hamiltonian constraint operator is obtained. Properties of the operator are studied by both analytical and numerical methods. The physical Hilbert space of the model is obtained by solving the quantum Hamiltonian constraint. An operator corresponding to the Dirac observable whose classical limit coincides with the mass of the BH is obtained. It is remarkable that the spectrum of this operator is discrete and does not contain zero. This result indicates the existence of a remnant after evaporation of BH.

The interior of a Schwarzschild BH can be foliated by spatially homogeneous 3-manifolds $\Sigma$ of topology $\mathbb R\times \mathbb S^2$.
We denote the natural coordinates adapted to the topology as $(x,\theta,\phi)$. In order to avoid the divergence of integration, we introduce a fiducial cell $\mathcal C\subset \Sigma$ with the same topology $\mathbb R\times \mathbb S^2$ but $x\in (0,L_0)$ and restrict all integrals in $\mathcal C$.  Because of the homogeneity, the classical phase space can be coordinatized by the canonical pairs $(b,p_b)$ and $(c,p_c)$. These symmetry-reduced variables are related to the Ashtekar-Barbero variables $(A_a^i,E_i^a)$ of the gravitational fields as \cite{ashtekar2005quantum}
\begin{equation}
\begin{aligned}
A_a^i\tau_i\dd x^a&=\frac{c}{L_0}\tau_3\dd x+b\tau_2\dd\theta-b\tau_1\sin\theta\dd\phi+\tau_3\cos\theta\dd\phi,\\
E_i^a\tau^i\partial_a&=p_c\tau_3\sin\theta\partial_x+\frac{p_b}{L_0}\tau_2\sin\theta\partial_\theta-\frac{p_b}{L_0}\tau_1\partial_\phi,
\end{aligned}
\end{equation}
where $\tau_j=-i\sigma_j/2$ ($j=1,2,3$) denote the basis of the Lie algebra $\mathfrak{su(2)}$ with $\sigma_j$ being the Pauli matrix. 
The non-vanishing Poisson brackets among the basic variables are
\begin{equation}
\begin{aligned}
\{c,p_c\}&=2G\gamma,\ \{b,p_b\}&=G\gamma,
\end{aligned}
\end{equation}
where $G$ is the gravitational constant and $\gamma$ is the Barbero-Immirzi parameter \cite{immirzi1993reality,barbero1994real}. 
The Hamiltonian constraint reads
\begin{equation}
C[N]=\frac{8\pi N}{\gamma^2}\frac{1}{|p_b|\sqrt{|p_c|}}   \left(\left(b^2+\gamma ^2\right) p_b^2+2 b p_b c p_c\right)=0.
\end{equation}
By choosing $N=|p_b|\sqrt{|p_c|}\gamma^2/(8\pi)$ which is proportional to the volume of $\mathcal C$, we get 
\begin{equation}\label{eq:clah}
C[N]=\h\equiv  2 p_b b c p_c+p_b^2b^2+\gamma ^2p_b^2=0.
\end{equation}
Note that, if one couples a massless scalar filed $\varphi$ to the model and use it to deparametrize the system, $\sqrt{\h}$ would be the physical Hamiltonian with respect to $\varphi$.

By loop quantization, one can get the kinematical Hilbert space $\tilde{\mathcal H}$ for the model of the Schwarzschild interior as
\begin{equation}
\tilde{\mathcal H}=\tilde\hmu\otimes\tilde\htau=L^2(\mathbb{R}_{\rm Bohr},\dd\mu_0)\otimes L^2(\mathbb{R}_{\rm Bohr},\dd\mu_0),
\end{equation}
where $\dd\mu_0$ is the Haar measure on the Bohr compactification $\mathbb{R}_{\rm Bohr}$ of $\mathbb{R}$ \cite{thiemann2008modern}. Let $|\mu\rangle\in \tilde\hmu$ and $|\tau\rangle\in\tilde\htau$ be the canonical basis which diagonalize the momentum operators $\hat p_b$ and $\hat p_c$. Then the holonomy operators $\widehat{e^{\pm i\delta_b b}}$ and $\widehat{e^{\pm i\delta_c c}}$, with some parameters $\delta_b$ and $\delta_c$, act on the basis as translations  respectively, i.e.,
\begin{equation}
\widehat{e^{\pm i\delta_b b}}|\mu\rangle=|\mu\pm 2\delta_b\rangle,\ \widehat{e^{\pm i\delta_c c}}|\tau\rangle=|\tau\pm 2\delta_c\rangle.
\end{equation}
The Hilbert space $\tilde{\mathcal H}$ is not separable. It is convenient to choose a separable subspace $\hil\subset\tilde\hil$ which is preserved by the basic operators $\hat p_b$, $\hat p_c$, $\widehat{e^{\pm i\delta_b b}}$ and $\widehat{e^{\pm i\delta_c c}}$ for some fixed $\delta_b$ and $\delta_c$ to study their properties. The solutions to the Hamiltonian constraint will be constructed through the separable subspace $\hil$. 
However, in the case of the black hole interior,  there are some ambiguities in choosing the parameter $\delta_b$ and $\delta_c$. Roughly speaking, the choices can be classified into three schemes. The first one is the so-called $\mu_o$-scheme where $\delta_b$ and $\delta_c$ are chosen to be constants \cite{ashtekar2005quantum,modesto2006loop}. The second one is the $\bar\mu$-scheme which allows $\delta_b$ and $\delta_c$ to be any functions of $p_b$ and $p_c$ \cite{boehmer2007loop,chiou2008phenomenological}. The third one is developed recently where $\delta_b$ and $\delta_c$ are phase space dependent only through Dirac observables \cite{corichi2016loop,ashtekar2018quantum}. 
The following analysis will be valid for the $\mu_o$-scheme, as well as for the special cases of the third scheme, where $\delta_c$ is constant but $\delta_b$ is a function of $p_c$ and $\sin(\delta_c c)$, e.g., 
$\delta_b=\sqrt{\Delta}/(2p_c\sin(\delta_c c))$ in \cite{corichi2016loop} with $\Delta$ being the area gap given by LQG. In the latter scheme, $\delta_b$ can also be treated as a c-number if $\hat \delta_b$ commutes with the studied operator. This is the case in our following analysis. Therefore, although $\delta_b$ is treated as a c-number, the results are still hold even if it is an operator as in the latter scheme.
Consider the separable Hilbert spaces $\hmuep\subset \tilde\hmu$ and $\htauep\subset\tilde\htau$, spanned by the bases $|\mu\rangle$ and $|\tau\rangle$ with $\mu=\varepsilon_b+2 n\delta_b$ and $\tau=\varepsilon_c+2 k\delta_c$  for $n,k\in \mathbb Z$  respectively. 
The operators that we are going to study will be restricted in some dense subspaces of $\hmuep$ and (or) $\htauep$.

To study the loop quantization of the Hamiltonian constraint, one needs to first regularize the classical expression corresponding to \eqref{eq:clah} in the full theory by the Thiemann's trick \cite{ashtekar2005quantum,corichi2016loop}, and then restrict the regularized expression into the symmetry-reduced model. This procedure leads to a regularized ``Hamiltonian constraint''
\begin{equation}\label{eq:reguclh}
\h_r=\frac{2}{\delta_b\delta_c}p_b\sin(\delta_b b)p_c\sin(\delta_c c)+\frac{1}{\delta_b^2}p_b^2\sin(\delta_b b)^2+\gamma^2p_b^2.
\end{equation}
Its constituents  $p_c\sin(\delta_c c)$ and $p_b\sin(\delta_b b)$ can be quantized respectively as 
\begin{equation}\label{eq:oppsin}
\begin{aligned}
\hat\beta_{\delta_b}&:=\frac{1}{2i}\left((\hpb+\gamma\Pl^2\b)\widehat{e^{i\delta_b b}}-\widehat{e^{-i\delta_b b}}(\hpb+\gamma\Pl^2\b)\right),\\
\hat\xi_{\delta_c}&:=\frac{1}{2i}\left((\hpc+\gamma\Pl^2\c) \widehat{e^{i\delta_c c}}-\widehat{e^{-i\delta_c c}}(\hpc+\gamma\Pl^2\c)\right),
\end{aligned}
\end{equation}
where $\Pl=\sqrt{G\hbar}$ is the Planck length,  $\b$ and $\c$ are some constants representing different operator-ordering strategies and their values can be calculated by considering the commutators $[\hat p_b,\widehat{e^{\pm i\delta_b b}}]$ and $[\hat p_c,\widehat{e^{\pm i\delta_c c}}]$. One can check that $\hat\xi_{\delta_c}$ is essentially self-adjoint, whose domain consists of the finite linear combinations of the basis $|\tau\rangle$ in $\htauep$. For a given $\delta_b$, $\hat \beta_{\delta_b}$ is also essentially self-adjoint with domain consisting of finite linear combinations of $|\mu\rangle$ in $\hmuep$. Their spectrum are both the entire real line. Moreover, they have both desirable classical limits. For instance, the action of $\hat\beta_{\delta_b}$ on $|\mu\rangle$
can be approximated by $-i\gamma\delta_b\Pl^2\sgn(\mu)\sqrt{|\mu|}\partial_\mu\sqrt{|\mu|}$ for $|\mu|\gg 1$. Hence $\hat\beta_{\delta_b}$ returns to the Schr\"odinger quantization of $ p_b b$ in this limit. Therefore, the classical limit of $\hat\beta_{\delta_b}$ is correct. Similarly, the classical limit of $\hat\xi_{\delta_c}/\delta_c$ corresponds to the variable $p_c c$, and $m:=cp_c/(L_0\gamma)=GM$ is a Dirac observable where $M$ is the ADM mass of the Schwarzschild BH \cite{ashtekar2018quantum}. Hence the property of the spectrum $m L_0\gamma \delta_c $
of $\hat\xi_{\delta_c}$ is the key issue in our following study, although there might be subtleties in defining Dirac observables in the framework of loop quantum Schwarzschild interior (see e.g. the arguments in \cite{bodendorfer2019mass}).

Now let us come back to the Hamiltonian constraint \eqref{eq:reguclh} itself. We obtain the corresponding operator as
\begin{equation}
\hp=\frac{2}{\delta_b\delta_c}\hat\beta_{\delta_b}\hat\xi_{\delta_c}+\frac{1}{\delta_b^2} \hat\beta_{\delta_b}^2+\gamma^2\hat p_b^2.
\end{equation}
Because $\hat\xi_{\delta_c}$ commutes with $\hp$, we can replace $\hat\xi_{\delta_c}$ by its eigenvalue $m L_0\gamma\delta_c$ to consider the operator
\begin{equation}\label{eq:hpm}
\hpm:=\frac{2m L_0\gamma}{\delta_b} \hat\beta_{\delta_b}+\frac{1}{\delta_b^2}\hat\beta_{\delta_b}^2+\gamma^2\hat p_b^2,
\end{equation}
defined in the Hilbert space $\hmuep$ for each given $\delta_b$. The operator $\hpm$ can be divided into the diagonal part and the off-diagonal part with respect to the canonical basis $|\mu\rangle$.  Then one can use the Kato-Rellich theorem in \cite{reed2003methods} to prove that  $\hpm$ is essentially self-adjoint with the domain consisting of finite linear combinations of $|\mu\rangle$. Moreover, one can prove that $\langle\psi|\hpm|\psi\rangle\geq -L_0^2m^2\gamma^2+\gamma^2\langle\psi|\hat p_b^2|\psi\rangle$ for all $|\psi\rangle$ in the domain of $\hpm$, and the operator $-L_0^2 m^2\gamma^2+\gamma^2\hat p_b^2$ in the right hand side has unbounded discrete spectrum. Thus the operator $\hpm$ has purely discrete spectrum according to the min-max theorem \cite{reed2003methodsvi}. Let us consider the perturbation of $\hpm$ with respect to $m$. Because $\delta_b$ could depend on $m$, the Hilbert space $\tilde{\hil}_{\b}^{\varepsilon_b}$ where $\hp^{\tilde{m}}$ is defined usually differs from the Hilbert space $\hmuep$ where $\hpm$ is defined. To compare the operator $\hp^{\tilde{m}}$ to $\hpm$, we can identify $\tilde{\hil}_{\b}^{\varepsilon_b}$ and $\hmuep$ by the unitary map $i:\tilde{\hil}_{\b}^{\varepsilon_b}\ni|\varepsilon_b+2n\tilde{\delta}_b\rangle\mapsto |\varepsilon_b+2n\delta_b\rangle\in\hmuep$. Then we may compare the operator $i \hp^{\tilde{m}} i^{-1}$ with $\hpm$. Suppose that $\delta_b$ is an analytic function of $m$ locally. This is the case in both schemes that we are considering. Then by defining a sequence of operators
$$
\hat{\mathfrak t}_n=\sum_{|\mu\rangle\in\hmuep }\frac{1}{n!}\frac{\dd^n \langle\mu|i^{-1}\hp^{\tilde{m}} i|\mu\rangle}{\dd \tilde m^n}\Big|_{\tilde m=m}|\mu\rangle\langle \mathfrak \mu|,
$$
one can show that for any $n\geq 1$,   
 there exist some positive numbers $a$ and $b$ such that
$
|\langle \psi |\hat{\mathfrak t}_n|\psi\rangle|\leq a \langle\psi | (\hpm+L_0^2m^2\gamma^2)|\psi\rangle +b\langle\psi|\psi\rangle
$
for all $|\psi\rangle$ in the domain of the operator $(\hpm+L_0^2m^2\gamma^2)^{1/2}$. Hence, the operators $\hp^{(m+\delta m)}$ for sufficient small $\delta m$ form a holomorphic family of type (B) in the sense of Kato \cite{kato2013perturbation}. Then each eigenvalue of $\hp^{(m+\delta m)}$ can be obtained through a perturbation around some eigenvalue $\omega_0$ of $\hpm$.  More precisely, if $\omega_0$ has the algebraic multiplicity $k$, 
$\hp^{(m+\delta m)}$ has exactly $k$ eigenvalues (counting multiplicity) near $\omega_0$. These eigenvalues are given by $p\ (\leq k)$ distinct, single-valued and analytic functions $\omega_1(\delta m),\cdots,\omega_p(\delta m)$
\cite{reed2003methodsvi,kato2013perturbation}.

To solve the Hamiltonian constraint, it is necessary to diagonalize the operator $\hpm$.
This can be realized by the approximation of finite-dimensional cut-off. Consider the finite-dimensional Hilbert spaces $\hmu^{(k)}$ spanned by the basis $|\mu\rangle\in\hmuep$ with $
|\mu|\leq 2 k \delta_b+\varepsilon_b$ for a positive integer $k$. Let  $\hpm_k$ be the restriction of $\hpm$ to $\hmu^{(k)}$.  Let $\lambda_n^{(k)}$ with $n\leq 2k+1$ be the $n$th eigenvalue of $\hpm_k$ and $\phi_n^{(k)}$ be a corresponding normalized eigenvector. The eigenvalues are ordered as 
 $\lambda_1^{(k)}\leq\cdots \lambda_n^{(k)}\leq\cdots\leq \lambda_{2k+1}^{(k)}$. Then it can be proven that  $\lim_{k\to \infty}\lambda_n^{(k)}=:\omega_n$ exists, and $\omega_n$ is the $n$th smallest eigenvalue of $\hpm$. Additionally, the weak limit point(s) of the sequence $\{\phi_n^{(k)}\}_{k\geq n}$ as $k\to\infty$ span the eigenspace of $\hpm$ corresponding to the eigenvalue $\omega_n$ for each $n$.
 Moreover, for $\omega_n\neq\omega_{n+1}$, we have $\omega_n<\omega_{n+1}$ and $
\sigma(\hpm)\cap(\omega_n,\omega_{n+1})=\emptyset$
where $\sigma(\hpm)$ denotes the spectrum of $\hpm$.  Therefore, for $n\ll k$, the $n$th smallest eigenvalue and the corresponding eigenvector of $\hpm$ can be well approximated by the $n$th smallest  eigenvalue and the corresponding eigenvector of $\hpm_k$. Let $
\varepsilon_n^{(k)}:=\|(\hpm-\lambda_n^{(k)})\phi_n^{(k)}\|$. Then the inequality
$\lambda_n^{(k)}-\varepsilon_n^{(k)}<\lambda_n<\lambda_n^{(k)}+\varepsilon_n^{(k)}$ can be used to control the errors of the numerical calculation.
A subtle issue of the numerical computation would be the choice of the constant $\b$ in \eqref{eq:oppsin} and the constant $\varepsilon_b$ used to define the Hilbert space $\hmuep$. 
However, the operator $\hpm$ with $\b\neq 0$ is different from the operator $\hpm$ with $\b=0$ by a small perturbation.  Moreover, one can identify $\hmuep$ for $\varepsilon_b\neq 0$ with the Hilbert space $\hmu^0:=\hmu^{\varepsilon_b=0}$ via the unitary map $\mathfrak{i}: \hmuep\ni |\varepsilon_b+2n\delta_b\rangle\mapsto |2n\delta_b\rangle\in\hmu^0$ and compare the operator $\hpm$ in $\hmu^0$ with the operator $\mathfrak{i}\hpm\mathfrak{i}^{-1}$ in which $\hpm$ is defined in $\hmuep$. The latter one differs from the former one by a small perturbation.  Hence we will choose $\b=0=\varepsilon_b$ for our computation without loss of generality.  
Then the action of $\hpm$ on a state $\psi(\eta)=\langle 2\delta_b\eta|\psi\rangle$ with $\eta\in\mathbb Z$ reads
\begin{equation}\label{eq:hexpand}
\begin{aligned}
&(\hpm\psi)(\eta)=-\frac{1}{4}\gamma^2\Pl^4(\eta+2)(\eta+1)\psi(\eta+2)\\
&-i\gamma^2\Pl^2L_0m(\eta+1)\psi(\eta+1)\\
&+\left(\frac{1}{4}\gamma^2\Pl^4(\eta+1)^2+\frac{1}{4}\gamma^2\Pl^4(1+4\delta_b^2\gamma^2)\eta^2\right)\psi(\eta)\\
&+i\gamma^2\Pl^2L_0m\eta\psi(\eta-1)-\frac{1}{4}\gamma^2\Pl^4\eta(\eta-1)\psi(\eta-2).
\end{aligned}
\end{equation}
For $\eta=0$, both the coefficients of the terms $\psi(-1)$ and $\psi(-2)$ vanish. Thus given an eigenvector $\psi(\eta)$ of $\hpm$, its values for $\eta<0$ decouple from its values for $\eta\geq 0$. Therefore, the eigenvector $\psi(\eta)$ can be classified into two supper-selected sectors. The first sector consists of those $\psi$ that are vanishing for $\eta<0$, while the second sector consists of those vanishing for  $\eta\geq 0$ \footnote{For $\varepsilon_n\neq 0$ or $\b\neq 0$, $\psi(\eta)=0$ becomes $\psi(\eta)\ll 1$  correspondingly for the two sectors. }. Let us denote the eigenvectors in the first and the second sectors as $\psi_+$ and $\psi_-$ respectively. Correspondingly, the eigenvalues will be denoted as $\omega^\pm$. It turns out that the state $(T\psi_-)(\eta):=\psi_-(-\eta-1)$ is an eigenvector of the operator $\hpm+\hat\epsilon$ corresponding to the same eigenvalue $\omega_-$, where $\hat\epsilon\psi(\eta)=-\gamma^4\Pl^4\delta_b^2(2\eta+1)\psi(\eta)$ is relatively very small with respect to $\hpm\psi(\eta)$.  Consequently,
for any given $\omega^+$ there is a unique adjoint $\omega^-$ nearby it, where $\omega^-$ is also an eigenvalue of the perturbed operator $\hpm+\hat\epsilon$. The corresponding eigenvectors of the two adjoint eigenvalues  satisfy
\begin{equation}\label{eq:sym}
\psi_+(\eta)\approx \psi_-(-\eta-1).
\end{equation}

\begin{figure}[!t]
\centering
\includegraphics[width=0.4\textwidth]{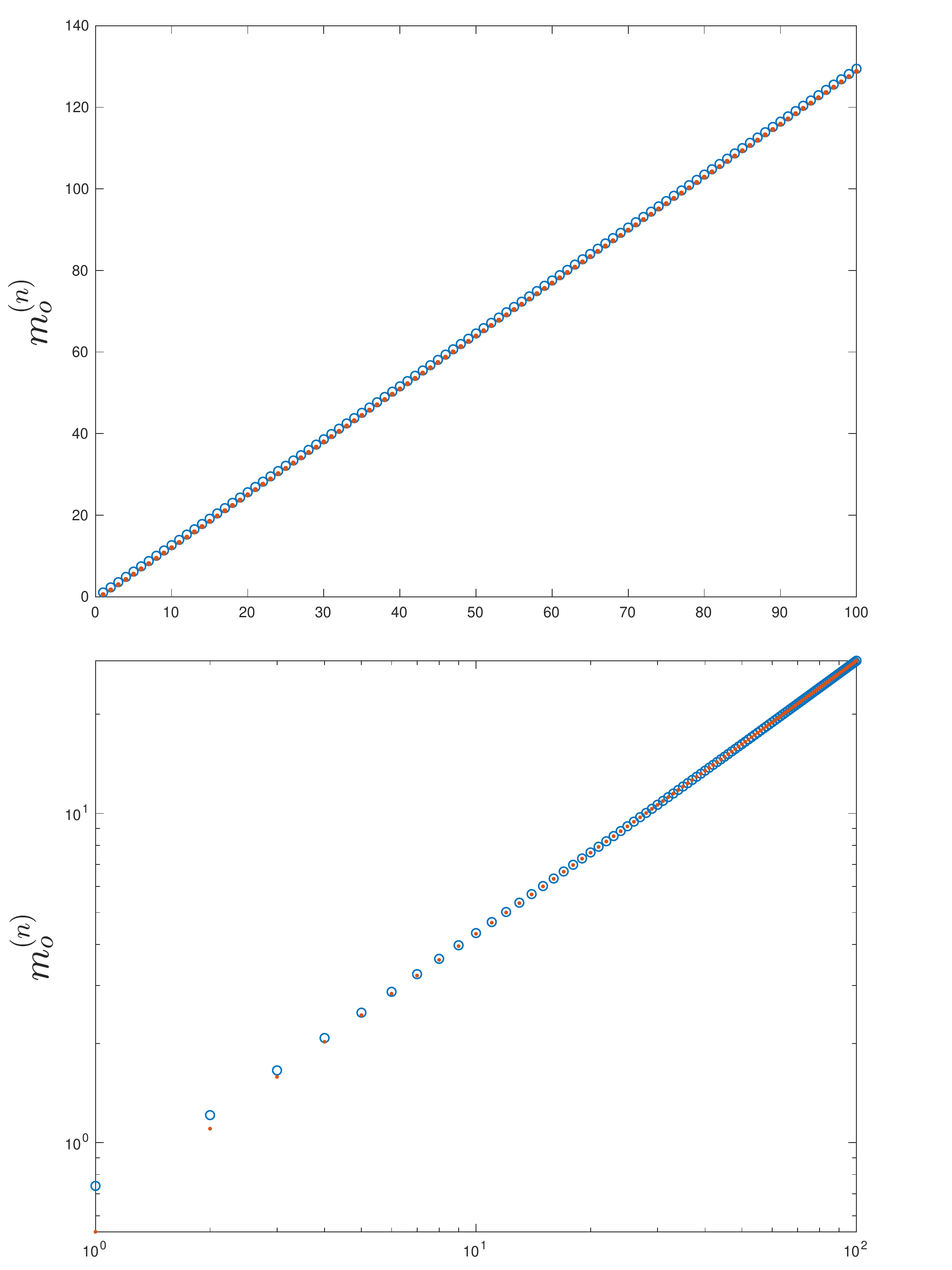}
\caption{Plots of $\sigma_\xi$ for the $\mu_o$ scheme with $\delta_b=\sqrt{\Delta}$ (top panel) and the scheme with $\delta_b=\sqrt{\Delta}/(2\,|m|)$ (bottom panel). As shown in the figure, nearby each value of $m_o^{(n)}$ there exists an adjoint value of $m_o^{(n')}$. $0$ does not belong to $\sigma_\xi$.}\label{fig:dism}
\end{figure}

Now let us solve the Hamiltonian constraint
\begin{equation}\label{eq:constraint}
\hp|\Psi\rangle=0.
\end{equation}
Any state $|\psi\rangle$ in the separable Hilbert space $\hil$ can be spanned as 
\begin{equation}
|\psi\rangle=\int_{\mathbb R}\dd m\sum_{\omega_m\in\sigma(\hpm)}\psi(m,\omega_m)|m,\omega_m\rangle,
\end{equation}
where $\sigma(\hpm)$ denotes the spectrum of $\hpm$, and $|m,\omega_m\rangle$ is the common eigenstate of $\hat\xi_{\delta_c}$ and $\hpm$ satisfying
\begin{equation*}
\begin{aligned}
\hat\xi_{\delta_c}|m,\omega_m\rangle&=m L_0\gamma\delta_c|m,\omega_m\rangle,\\
\hpm|m,\omega_m\rangle&=\omega_m|m,\omega_m\rangle.
\end{aligned}
\end{equation*}
 The eigenstates are normalized as
 \begin{equation*}
 \langle m',\omega_{m'}|m,\omega_m\rangle=\delta(m,m')\delta_{\omega_m,\omega_{m'}},
 \end{equation*}
where $\delta(m,m')$ is the Dirac $\delta$-distribution and $\delta_{\omega_{m},\omega_{m'}}$ is the Kronecker delta. Then the action of $\hp$ on $|\psi\rangle$ is given by 
\begin{equation}
\hp|\psi\rangle= \int_{\mathbb R}\dd m\sum_{\omega_m\in\sigma(\hpm)}\omega_m\psi(m,\omega_m)|m,\omega_m\rangle.
\end{equation} 
Thus the solutions to Eq. \eqref{eq:constraint} take the form
\begin{equation}\label{eq:solution}
|\Psi\rangle=\int_{\mathbb R}\dd m\sum_{\omega_m\in\sigma(\hpm)}\delta_{\omega_m,0}\,\psi(m,\omega_m)|m,\omega_m\rangle.
\end{equation}
Because of the summation over $\sigma(\hpm)$ and the factor $\delta_{\omega_m,0}$, we only need to consider those $m$ such that 
$0\in\sigma(\hpm)$. Suppose $m_o$ satisfies $0\in\sigma(\hp^{(m_o)})$. According to the properties of $\hpm$, the eigenvalues of $\hp^{(m_o+\delta m)}$ for $\delta m\ll 1$ can be expanded at $m_o$ by a power series of $\delta m$. Together with the fact that $\sigma(\hp^{(m_o)})$ is discrete, the eigenvalues of $\hp^{(m+\delta m)}$ are in general not vanishing. That is, $0$ usually may not belong to $\sigma(\hpm)$. Thus it is reasonable to expect that there are only countably many $m$  such that $0\in\sigma(\hpm)$. This speculation is confirmed by our numerical computation in the $\mu_o$-scheme as well as the scheme with $\delta_b=\sqrt{\Delta}/(2\,|m|)$, as shown in Fig. \ref{fig:dism}. 
Denote each of these $m$ as $m_o^{(n)}$ with a $n\in\mathbb Z$ and the set of $m_o^{(n)}$ as $\sigma_\xi$.  
 Then the term $\delta_{\omega_m,0}\psi(m,\omega_m)$ in \eqref{eq:solution} would vanish for $m\notin \sigma_\xi$. Thus the resulted solutions of \eqref{eq:solution} will have vanishing norm for any regular functions $\psi(m,\omega)$. Therefore,
  $\psi(m,\omega_m)$ have to be chosen as
   the Dirac $\delta$-distributions.
   Then the solutions to  Eq. \eqref{eq:solution} can be written as
 \begin{equation}\label{eq:projection}
 |\Psi)=\int_{\mathbb R}\dd m\sum_{n}\delta(m_o^{(n)},m) \psi(m,0)|m,0\rangle.
 \end{equation}
 These $|\Psi)$ are anti-linear functionals on a dense subspace
  $\mathcal S\subset\hil$, which supports the Dirac $\delta$-distributions.
  Eq. \eqref{eq:projection} gives a projection map
 $$
  \mathbb P=\int_{\mathbb{R}}\dd m\sum_n\delta(m,m_o^{(n)})|m,0\rangle\langle m,0|
  $$ from $\mathcal{S}$ to the solution space. Therefore, by the refined algebra quantization procedure \cite{thiemann2008modern}, the physical inner product of two solutions  reads
  \begin{equation}
  (\Psi_1|\Psi_2)=\langle \psi_1|\mathbb{P}|\psi_2\rangle=\sum_{n\in\mathbb Z}\overline{\psi_1(m_o^{(n)})}\psi_2(m_o^{(n)}),
  \end{equation}
  where the line over a function denotes its complex conjugation. 
  Thus the physical Hilbert space of the solutions is isometric to the Hilbert space
\begin{equation}
\hbh:=\overline{\{f:\sigma_\xi\to \mathbb C,\ \sum_n|f(m_o^{(n)})|^2<\infty\}},
\end{equation}
where the line over a space denotes its completion. 
The Dirac observable $\hat\xi_{\delta_c}$ in $\hil$ can be promoted to an operator $\hat \xi'_{\delta_c}$ in $\hbh$ by the dual action, which gives
\begin{equation}
(\hat \xi_{\delta_c}'f)(m_o^{(n)})=L_0\gamma \delta_c m_o^{(n)}f(m_o^{(n)}),\ \forall\ m_o^{(n)}.
\end{equation}
This formula implies that each $m_o^{(n)}$ is an eigenvalue of $\xi_{\delta_c}'$, and $\xi_{\delta_c}'$ is self-adjoint in $\hbh$ with the spectrum $\overline{\sigma_\xi}$ as the closure of $\sigma_\xi$. 
Let us consider the properties  of $\overline{\sigma_\xi}$. Firstly, the identity  $\overline{\hpm\psi}=\hp^{(-m)}\overline\psi$ ensures that $-m_o^{(n)}\in\overline{\sigma_\xi}$ if $m_o^{(n)}\in\overline{\sigma_\xi}$. 
Secondly, it is easy to check that there is no nontrivial state $\phi\in\hmuep$ such that 
$
\hat\beta_{\delta_b}\phi=0=\hat p_b\phi.
$
Hence, Eq.\eqref{eq:hpm} implies that $0\notin \sigma(\hp^{(0)})$. Moreover, 
thanks to the holomorphicity of $\hpm$, for sufficient small $\delta m$, each eigenvalue of $\hp^{(\delta m)}$ can be obtained through an analytic perturbation of some eigenvalue of $\hp^{(0)}$. 
This ensures $0\notin \sigma(\hp^{(\delta m)})$  for all of these $\delta m$. Therefore, we conclude that there exists a gap between the spectrum $\overline{\sigma_\xi}$ and $0$, i.e., $0\notin \overline{\sigma_\xi}$.   

The numerical result in Fig. \ref{fig:dism} shows that the values of $m_o^{(n)}$ are discrete and have the following characters. First, for each $m_o^{(n)}$, there exists an adjoint $m_o^{(n')}$ nearby it. This property comes from the symmetry \eqref{eq:sym} of the eigenvectors of $\hpm$. Second, the lowest value of $|m_o^{(n)}|$ in $\sigma_\xi$ turns out to be: $|m_o^{(\rm lwt)}|=0.5499$ for the $\mu_o$ scheme and  $|m_o^{(\rm lwt)}|=0.5362$ for the other scheme.

%

In summary, the interior of the Schwarzschild BH was quantized by LQG method. By studying the properties of the Hamiltonian constraint operator $\hp$ in details,  the physical Hilbert space $\hbh$ describing the Schwarzschild interior was obtained. 
The spectrum $\overline{\sigma_\xi}$ of the Dirac observable $\hat\xi_{\delta_c}'$ in $\hbh$ was analyzed by both analytical and numerical methods. It turns out that the $\overline{\sigma_\xi}$ is discrete and $0$ is not contained in $\overline{\sigma_\xi}$, i.e., there exists a gap between $\overline{\sigma_\xi}$ and $0$ in the schemes that we were considering. Since the classical limit of $\hat\xi_{\delta_c}'$ is proportional to the ADM mass of the Schwarzschild BH, our result indicates the existence of the BH remnants after evaporation. Moreover, this conclusion holds for all the schemes such that the length parameter $\delta_b$ of the holonomy operator is a locally analytic function of the spectrum parameter $m$ of the kinematical correspondence $\hat\xi_{\delta_c}$ of the Dirac observable $\hat\xi_{\delta_c}'$.  The BH remnants  predicted by our LQG model
 lay a theoretical foundation to consider them as dark matter candidates, as well as to solve the puzzle of information loss in BH evaporation.  Moreover, it is possible to use
 the numerical method developed in this paper to further study the properties of the BH remnant, the back reaction of the Hawking radiation and the distortion of the Hawking spectrum resulted from the quantum gravity effects.

This work is supported by NSFC with Grants No. 11875006, No. 11961131013 and No. 11775082.  CZ acknowledges the support by the Polish Narodowe Centrum Nauki, Grant No. 2018/30/Q/ST2/00811.



%


\end{document}